\documentclass[12pt]{article}
\usepackage{amssymb}
\usepackage{amsfonts}
\usepackage{graphicx}

\begin{document}

\title{{\large \textbf{The evolution of Complexity co-occurring keywords: bibliometric analysis and network approach}}}
\author{{\small Tanya Ara\'{u}jo$^{a,b}$ and Alexandre Abreu$^{a,c}$ and  Francisco Lou\c{c}\~{a}$^{a,b}$} \and {\small $^{a}$ISEG
(Lisbon School of Economics \& Management) Universidade de Lisboa, }
\and {\small $^{b}$Research Unit on Complexity and Economics (UECE) and  \small $^c$ CEsA/ISEG, Portugal}}

\date{}
\maketitle

\abstract{Bibliometric studies based on the Web of Science (WOS) database have become an increasingly popular method for analysing the structure of scientific research. So do network approaches, which, based on empirical data, make it possible to characterize the emergence of topological structures over time and across multiple research areas. Our paper is a contribution to interweaving these two lines of research that have progressed in separate ways but whose common applications have been increasingly more frequent.
Among other attributes, Author Keywords and Keywords Plus® are used as units of analysis that enable us to identify changes in the topics of interest and related bibliography. By considering the co-occurrence of those keywords with the Author Keyword \texttt{Complexity}, we provide an overview of the evolution of studies on Complexity Sciences, and compare this evolution in seven scientific fields.
The results show a considerable increase in the number of papers dealing with complexity, as well as a general tendency across different disciplines for this literature to move from a more foundational, general and conceptual to a more applied and specific set of co-occurring keywords. Moreover, we provide evidence of changing topologies of networks of co-occurring keywords, which are described through the computation of some topological coefficients. In so doing, we emphasize the distinguishing structures that characterize the networks of the seven research areas.
}

\vspace{0.5cm}

\textbf{keywords:} Complex systems, Complexity, Bibliometrics, Scientometrics, author Keywords, Keywords Plus

\vspace{0.25cm}


\maketitle
\newpage
\section{Introduction}
This paper provides an overview of the evolution of studies on complexity, comparing the framework of seven social and natural scientific fields and noting, as time goes by, the change of topics of interest, analytical techniques and even scientific priorities. After indicating some relevant compared literature, we present our data and bibliometric methods, then we proceed to detect evidence of changing topologies of networking and, finally, some conclusions are presented. It is noticeable that the very concept of complexity changed over time.

In the following, we will show evidence of the evolution and enlargement of the concepts of complexity in different sciences, which is consistent through time and itself an adaptive process evidencing the role and contribution of more realistic hypotheses on nature than those of general equilibrating systems.

\subsection{Some compared literature}

In previous reviews for the case of economics, such as that of reference \cite{Dur2012}, complexity approaches are reduced to "deviations from the baseline general equilibrium model of an economy under uncertainty" and not to challenges to it. The author defines a complex system as "a system comprised of a population of interacting heterogeneous agents in which the behavior of each agent can be described as a function of the behavior of other agents, as well as other factors", which is a fair description of the main characteristics of the approach. Nevertheless, Durlauf stands by the rejection of suggestions that this is a new paradigm, only accepting its constitutional contradiction with the description of nature based on a representative agent, which, by the way, constituted for long the basis for the vindication of a micro-foundation for economics and a description of a model of rationality. Yet, the general properties of heterogeneity of beliefs, experience, behavior including learning and strategy, as well as interdependence and conflict, institutions and coordination, and therefore emergence of new properties or phase transition, cannot be all found in neoclassical models or be reduced to deviations from it. Indeed, this approach frequently leads to the abandon of methodological individualism. This may explain Durlauf's hostility to econophysics and, in particular, to Physica A, as an "unhealthy development" and "diletanttism".

Other authors have been more sympathetic towards the incorporation of elements from the complexity approach in economics, while not necessarily identifying any radical paradigmatic incompatibility. Reference \cite{Holt2011}, for example, argued just over a decade ago that what we were already upon an "era of complexity" in economics, with complexity economics "becoming just economics", as part of the process of economics in general becoming more empirical and less deductive. A convergent view was put forth around the same time by reference \cite{Coop2011}, who identified two traditions in economic thought as constituting a "prehistory of complexity" in economics, and argued that the 2008 financial crisis created an environment more attuned to complexity, especially in the study of finance and financial regulation.

Recent systematic reviews have certainly identified a growing literature on complexity within economics, or which adopt a complex systems perspective to the analysis of specific topics in economics. Reference \cite{Ale2022} identified 523 articles on Google Scholar by using a search strategy that combined economics and complexity-related terms. They subsequently narrowed this sample down to 113 core articles using a systematic review and meta-analysis protocol, based on which they categorised recent research as including macro, micro and meso applications. These authors share the view that the aftermath of the financial crisis led to academic scholarship "further question[ing] the use of traditional ‘linear’ economic ontologies". In their turn,  reference \cite{Zho2022} was able to locate 1302 articles in the Web of Science database published between 1992 and 2021 dealing with Agent-Based Models in finance. Using bibliometric coupling methods, they identified two main clusters of scholars within this subfield: the econophysics research community and the economics and finance community, each of which publishes in its own set of journals and mostly cites other authors within the same research community without crossing over to the other.

Alongside the intellectual and scholarly repercussions of the financial crisis, at least two other developments have contributed significantly to the burgeoning volume of the literature on complexity in economics. The first one is the increasing amount of work on the economics of climate change, which is especially consistent with the adoption of a complexity approach. In a non-systematic review published seven years ago,  reference \cite{Bal2017} surveyed papers dealing with the micro- and macroeconomics of climate change using a Complexity Sciences perspective, and concluded that such a perspective has yielded important insights in four major areas: "i) coalition formation and climate negotiations; ii) macroeconomic impacts of climate-related events; (iii) energy markets and (iv) diffusion of climate-friendly technologies". The other development is the work on "economic complexity" in the rather different sense of productive differentiation and sophistication, as developed over the last two decades by scholars such as Cesar Hidalgo \cite{Hid2021}, and which has led, inter alia, to the establishment of the Observatory of Economic Complexity and the creation of the Economic Complexity Index. It is worth pointing out that while this strand of the economic literature is characterised by sophisticated empirical methods, it seems to have little in common with the complex systems approach in economics as traditionally understood - even though the shared terminology has inevitably contributed to the rapidly increasing bibliometry, and possibly to some confusion around the meaning in each case of such terms as "economic complexity" or "product complexity".

Other bibliometric research has concentrated on notions of complexity in different fields. A study on social cohesion \cite{Mous2022} took a sample of 5027 articles from 2362 journals for the period 1994-2020, investigating how processes such as migration, ageing or the structural evolution of economies, with a wide range of dimensions and implications, could be assessed from the cooperation among several disciplines. The author proceeded to the definition of search terms, choice of databases and network analysis using tools as ours, although applied to a small number of contributions. Reference \cite{Sua2022} chose 428 articles from Scopus and the period 1937 to 2022 to detect networks in education studies, using co-occurrence analysis. They were interested in the assessment and management of uncertainty in education, and the evolution of related disciplines was required for that.  Reference \cite{Chu2019} focused on healthcare and chose 2505 articles from 268 journals, although only 454 articles were discussed in detail. These authors detect an evolution from conceptual to concrete papers, in particular in the use of the notion of complex adaptive systems as applied to healthcare.

 Reference \cite{Swe2020} proposed a first contribution on the bibliometry of the relation between climate change and epidemiology, namely on the health risks with an emphasis on infectious diseases, since there is evidence of increased survival rates and transmission of pathogens or vectors of agents causing diseases, including vulnerability to antimicrobial resistance, given global warming. The author studied 4247 document from Scopus and the period 1980-2019, detecting a pattern of new interest since 2007. In each of these cases, a small number of keywords was used, and dates of publication, analysis of content and of citations, and co-authoring were considered in network analysis. Reference\cite{Fracc2018} undertook a systematic review of articles on complex systems and resilience, followed by a co-authorship and cross-citation network analysis. From an original pool of 458 publications between 1997-2017 from the Web of Science database, they retained 154 papers deemed most relevant, and concluded that research on resilience and complexity has been carried out in a variety of disciplines, but in a largely fragmented way, with little cross-fertilisation, by relatively isolated groups.

Both the increase in the number of papers dealing with complex systems and their general tendency across different disciplines for this literature to move from the more foundational, general and conceptual to the more applied, specific and empirical are at least partly explained by methodological and technological developments in data science, artificial intelligence and machine learning. This is a point made in several different review papers in different fields. For example,  reference \cite{Maa2017} argues that that is currently the case in complex systems biology, while  reference \cite{Holo2017} make the case that the "availability of more and more data in disciplines and fields beyond physics", from cities to the dynamics of societies to textual data, makes it increasingly possible to apply the methods and insights of "physics beyond physics".

In ecology,  reference \cite{Maa2017} similarly argue that progress in Agent-Based Modelling was for some time slower than anticipated due to unanticipated technical difficulties, but that there is now an explosion of work along these lines, which is  accounted for by developments in data availability and machine learning technologies. And in another example, in this case from computer science, \cite{Xiao2015} review developments in complex system computation and their application to complex engineering problems, and argue that emergent computation based on decentralised and parallel modules collaborating with one another is currently "the forefront of complex science".

Meanwhile, this growth spurt in the applied literature has not precluded developments in more theoretical and abstract aspects of complex systems from continuing to take place. For example, an emerging literature strand in physics and mathematics, as exemplified by  references \cite{Lamb2019} and \cite{Bat2021}, proposes to develop the modelling of complex systems by going beyond network models characterized by pairwise interactions and instead modelling higher-order interactions involving groups of larger number of units, which are arguably better able to reproduce the dynamic properties of many complex systems.  Reference\cite{Liu2016}, also in a physics context, review advances with respect to the specific question of the control of complex systems, and conclude that the depth and breadth of applications of this field is likely to spawn many research communities in the next decade. In their turn,  reference  \cite{Pei2019} survey a large number of  algorithms for identifying key influencers (such as opinion leaders, epidemic superspreaders or keystone species) and discuss methods for locating essential nodes and dismantling networks, but conclude that the real-world applications of these algorithms are still limited due to the mismatch between the ideal conditions for which they were developed and the noise and errors that characterise the real world.

Brian Arthur \cite{Arthur2021} presented a recent overview of the field of complexity economics, described as a movement within economics and not as a theory per se. Indeed, Arthur, who coined the term "complexity economics" in 1999, notes that it has been a challenge to traditional neoclassical economics, which is based on the unrealistic assumptions of equilibrium plus rationality and thus defining well shaped problems of optimization. If instead economics is concerned with interaction among imperfect information of agents, reacting with their experience and expectations to change, then adaptation, strategies and beliefs become the central issues. Arthur emphasizes the diversity of approaches (neural networks, artificial intelligence, behavioral studies) and tools (nonlinear stochastic processes, econophysics, agent-based modeling, research on institutions) that are part of this field. The key method, agent-based modeling, has been largely developed in recent years, as our paper confirms.

In the following sections, we seek to make sense of the research on complexity in a variety of different fields as it has evolved over the last two decades, by considering the entire scientific production associated with complexity and complex systems in the Science Citation Index Expanded and Social Sciences Citation Index editions of the Web of Science database.

\subsection{Research questions}

The main questions addressed in this paper are the following:

\begin{enumerate}
    \item How different is the notion of Complexity when used as a keyword in different research areas and in two distant time periods?
       \item Can Author Keywords that co-occur with the keyword \texttt{Complexity} reveal important differences across multiple research areas?
          \item Would the analysis of Keywords Plus® contribute to improve the answers to the above questions?
          \item How different are the topological structures of the co-occurrence networks of Author Keywords across seven research areas?
\end{enumerate}

The next section presents the data collected from the Web of Science in May 2023. Section three presents and analyses the distributions of the frequency of keywords co-occurring with \texttt{Complexity} in different research areas and in two distant time periods. In Section four, the same analysis is performed by considering the frequency of Keywords Plus®. Section five describes the method used in the definition of networks of Author Keywords, their corresponding minimum-spanning-trees (MST) and the resulting network structures obtained for the research areas. The last section concludes.

\section{Web of Science (WoS) Data}

 Our bibliometric research has concentrated on notions of complexity in different research areas by considering the scientific production associated with complexity and complex systems in the Science Citation Index Expanded and Social Sciences Citation Index editions of the Web of Science database.

\begin{itemize}
    \item \textbf{Database}:{ Web of Science Core Collection}
    \item \textbf{Editions}:
    \begin{itemize}
    \item Science Citation Index Expanded (SCI)
    \item Social Sciences Citation Index (SSCI)
    \end{itemize}
    \item \textbf{Publication Years}:
     \begin{itemize}
    \item from 2000 to 2004
    \item from 2019 to 2023
    \end{itemize}
 \item \textbf{Document Type}: Article
    \begin{itemize}
       \item {Results for 2000-2004: }\textbf{4.000.000 articles}
        \item {Results for 2019-2023: }\textbf{10.000.000 articles}
    \end{itemize}
   \item \textbf{{Research Areas}}
    \begin{enumerate}
     \item Computer Science
     \item Mathematics
     \item Economics (and Management and Business and Finance)
    \item Physics
    \item Biology
     \item Sociology (and Social Sciences)    \item Neurosciences

   \end{enumerate}
    \end{itemize}

\begin{figure}[h!]
  \centering
\includegraphics[width=.55\textwidth]{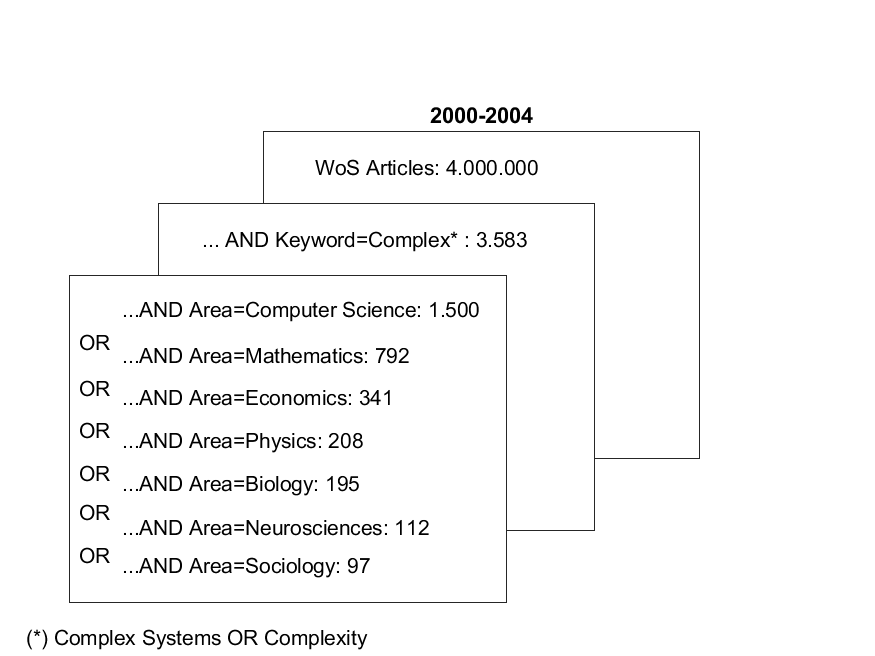}\includegraphics[width=.55\textwidth]{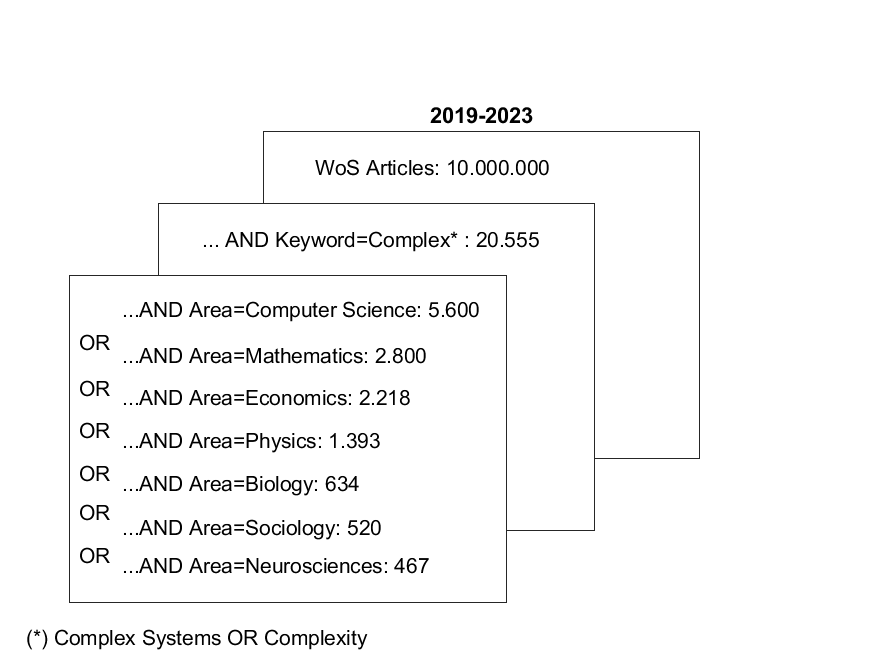}
  \caption{Number of papers in each research area for the two different time intervals.}
  \label{fig1}
  \end{figure}

Fig.\ref{fig1} shows, for each time period 2000-2004 (left) and 2019-2023 (right), the number of papers with the Author Keyword  \texttt{Complexity} by research area. There is a large increase in the number of papers from the first to the second time interval: 4.000.000 to 10.000.000. And an even larger increase is observed in the number of papers having \texttt{Complexity} as an Author Keyword.

\begin{figure}[h!]
  \centering
\includegraphics[width=.5\textwidth]{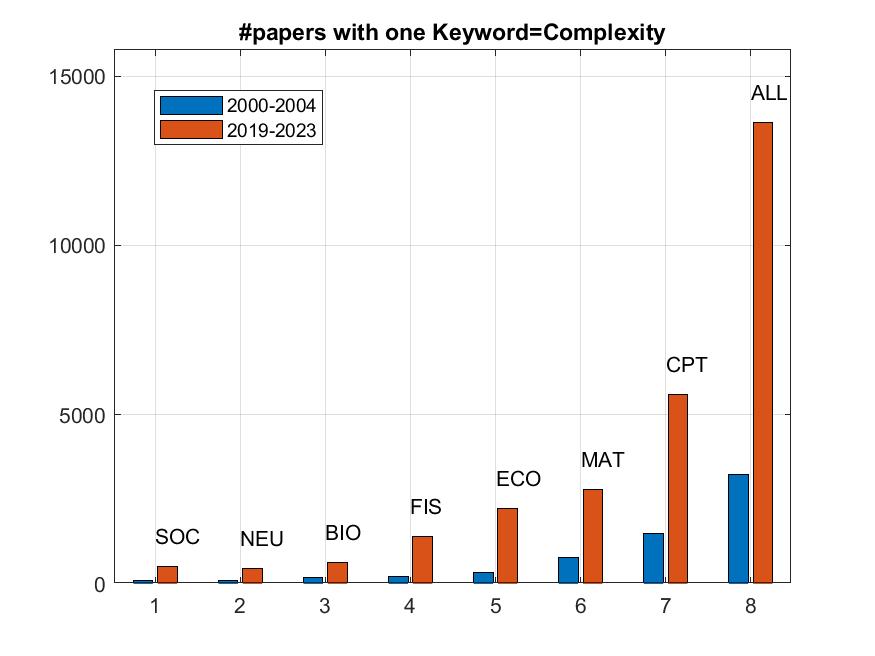}\includegraphics[width=.5\textwidth]{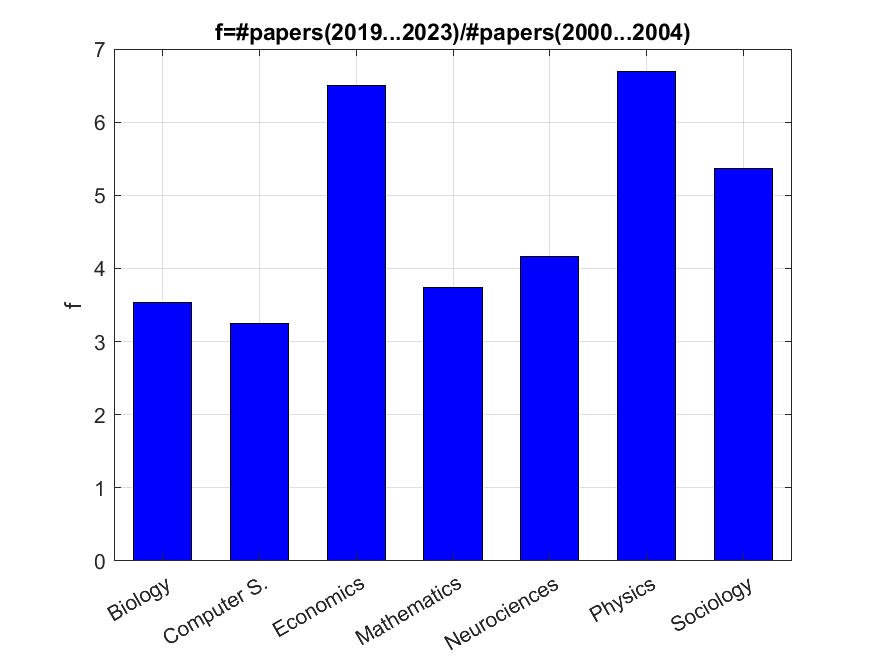}
  \caption{(a) The distribution of the number of papers with the Author Keyword \texttt{Complexity} along the research areas for the two time intervals (b) The distribution along the seven research areas of the ratio $f$ between the number of papers with the Author Keyword   \texttt{Complexity} in the second period (2019-2023) and the first period (2000-2004)}
  \label{fig2}
  \end{figure}

The first histogram $(a)$ in Fig.\ref{fig2} shows, for each research area and time period, the distributions of the number of papers with the Author Keyword  \texttt{Complexity}. The second histogram $(b)$ in Fig.\ref{fig2} shows the distribution of the ratio $f=\frac{\#papers(2019-2023)}{\#papers(2000-2004)}$ of the number of papers with the Author Keyword \texttt{Complexity} in the first time interval (2000-2004) and the number of papers with the Author Keyword  \texttt{Complexity} in the second one (2019-2023).

 The values of the ratio $f$ in the second histogram of Fig.\ref{fig2} show that the research areas of Physics ($f$=6.5) and Economics ($f$=6.25) are those displaying the largest increase in time of the number of papers with the Author Keyword \texttt{Complexity}. Physics and Economics are followed by Sociology, with a ratio of 5.2. The average value of the ratio computed for the seven areas (ALL) in the first histogram $(a)$ of Fig.\ref{fig2} is around four.

\section{Co-occurring Author Keywords}

The number of occurrences of an Author Keyword
is defined as the frequency of a keyword in the set of papers retrieved by research area. Additionally, and because we are interested in  the Author Keywords that co-occur with \texttt{Complexity}
the co-occurrence frequency is defined as the frequency of each keyword occurring simultaneously with the keyword \texttt{Complexity} by research area and time period.

The next two figures show, for each research area and time period (2000-2004 or 2019-2023), the Author Keywords  that co-occur with \texttt{Complexity}. Black bars represent the number of co-occurring Author Keywords  in the first time interval, while white bars represent those in the second one. The amounts of overlap  of co-occurring Author Keywords  between the two time intervals is represented by overlapping white and black bars, being quantified in the $y-$label. Although the histograms show just the most frequent 15 Author Keywords in each period, in the calculation of overlap, such a restriction in the number of Author Keywords being considered does not apply.

Overlap is computed as in reference \cite{Zha2016}.

\begin{equation}
    O=100\frac{Old(v_i) \cap New(v_j)}{Old(v_i) \cup New(v_j)}
\end{equation}

where $Old(v_i)$ and $New(v_j)$ are the sets of co-occurring Author Keywords  in the first (2000-2004) and the last (2019-2023) time intervals, respectively.

\begin{figure}[h!]
  \centering
\includegraphics[width=.5\textwidth]{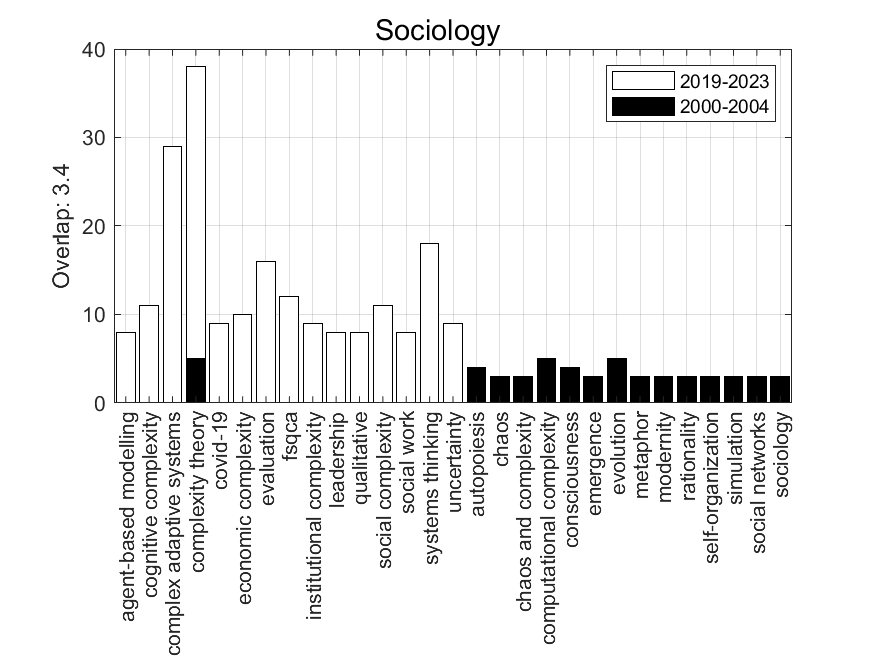}\includegraphics[width=.5\textwidth]{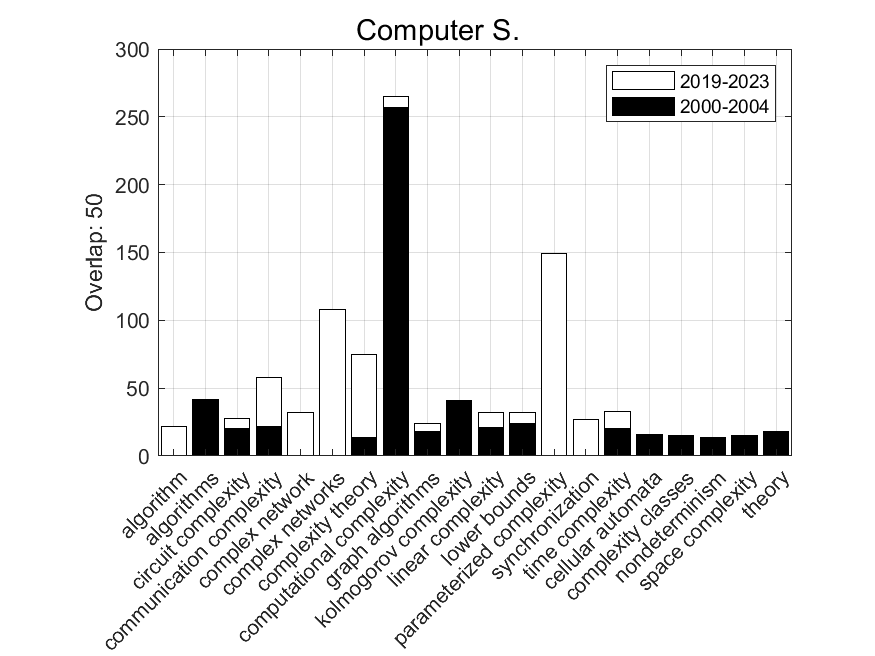}
\includegraphics[width=.5\textwidth]{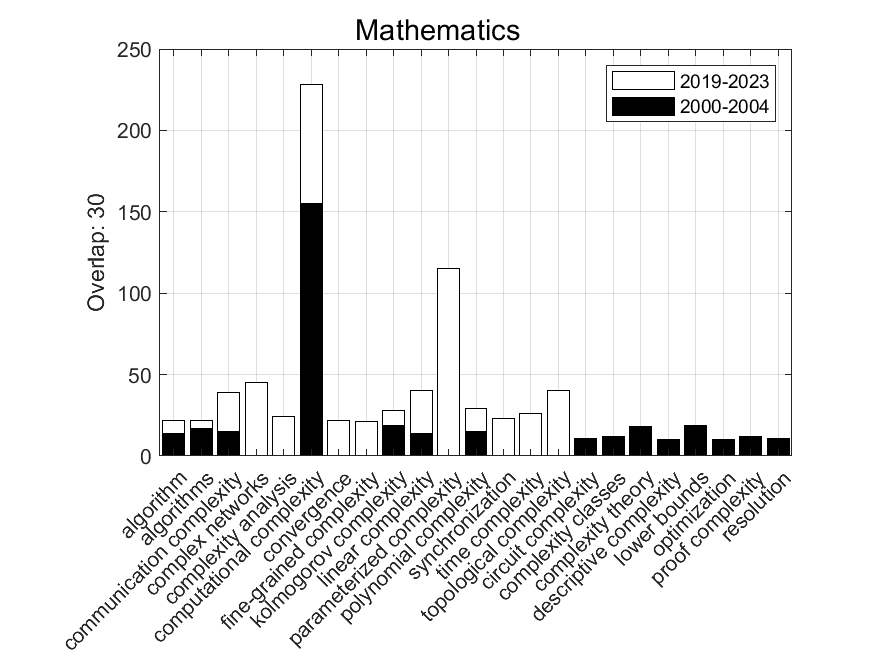}\includegraphics[width=.5\textwidth]{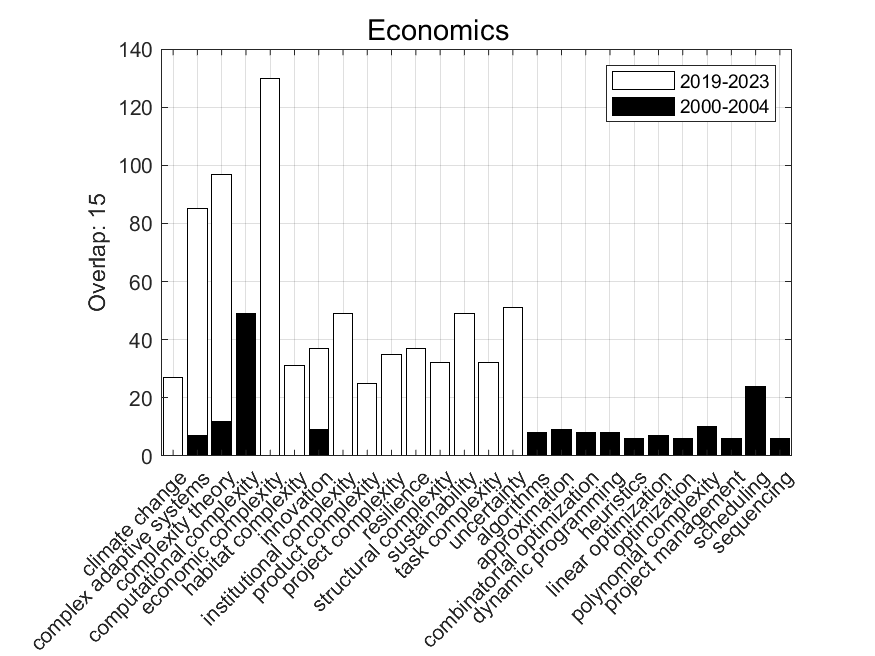}
  \caption{Author Keywords that co-occur with \texttt{Complexity}: in black in the first time interval (2000-2004) and colored white in the second one (2019-2023).}
  \label{fig3}
  \end{figure}

\begin{table}[b!]
    \centering
    \begin{tabular}{ccc}
    \textbf{Foundational} & \textbf{Tool} & \textbf{Specific} \\
    \hline
    Chaos & ABM & Climate Change\\
    Fractals & Networks & Habitat Complexity\\
    Self-Organization & Stochastic Process & Financial Market\\
    Emergence & Algorithms & Aging\\
    Entropy & Volatility & Sustainability\\
    Path dependence & Dimensions & EEG\\
    Autopoiesis & Simulations & Social work\\
    Self-similarity & Power laws & Working memory\\
    Complex adaptive systems & Matching models & Biodiversity\\
    \hline
\end{tabular}
\caption{A tentative classification of some Author Keywords co-occurring with \texttt{Complexity}}
\label{tab1}
\end{table}

Author Keywords  co-occurring with \texttt{Complexity} may be split in three categories: foundational concepts, interdisciplinary tools and specific concepts. Tab.\ref{tab1} comprises a tentative classification of some examples of Author Keywords co-occurring with \texttt{Complexity} in the seven research areas and helps to highlight some pieces of evidence from the analysis of the Author Keywords.

Author Keywords  co-occurring with \texttt{Complexity} in the earlier time period (2000-2004) tend to rely mostly on foundational concepts like \texttt{Fractal Dimension, Self-Organization, Emergence, Stochastic Process, Entropy} and \texttt{Chaos}.

 Author Keywords classified as interdisciplinary tools include: \texttt{Networks}, \texttt{Algorithms}, \texttt{Correlation Dimension} and \texttt{Agent-based Models (ABM)}. They appear in both time periods, being dependent on the research area whether they are found in the first or in the last time period. Unlike the foundational concepts, their association with a specific time period is less noticed.

 Author Keywords  co-occurring with \texttt{Complexity} in the later time period (2019-2023) are more likely to belong to the specific research area under study. Examples are the keywords \texttt{Convergence} in Mathematics, \texttt{Celular Automata} in Physics, \texttt{Sustainability} in Economics, \texttt{Biodiversity} in Biology, \texttt{Aging} in Neurosciences and \texttt{Leadership} in Sociology.

 In Sociology, (first histogram in Fig.\ref{fig3}) shows that \texttt{Complexity Theory} is the single occurring Author Keywords  that overlaps in time. Interestingly, we observe that in the first period 2000-2004, the authors in Sociology chose co-occurring Author Keywords  amongst those typical and foundational ones in Complexity Sciences, such as \texttt{Chaos, Emergence, Simulation}  and \texttt{Self-organization}. Later, such a choice shows an important shift towards \texttt{Agent-based modelling}, \texttt{Computational Complexity} and \texttt{Complex adaptive systems}, which are also non-specific sociological concepts.

The second histogram in Fig.\ref{fig3} shows that, as expected, Computer Science displays the largest overlap of co-occurring Author Keywords  in the two time periods. This area is, naturally, followed by Mathematics. In both areas, \texttt{Computational Complexity} is the leading co-occurring Author Keywords   found in both the first and the second time period. \texttt{Computational Complexity} is followed by \texttt{Algorithms}, \texttt{Kolmogorov Complexity} and \texttt{Linear Complexity}, both in Computer Science and Mathematics.
\begin{figure}[h!]
  \centering
\includegraphics[width=.5\textwidth]{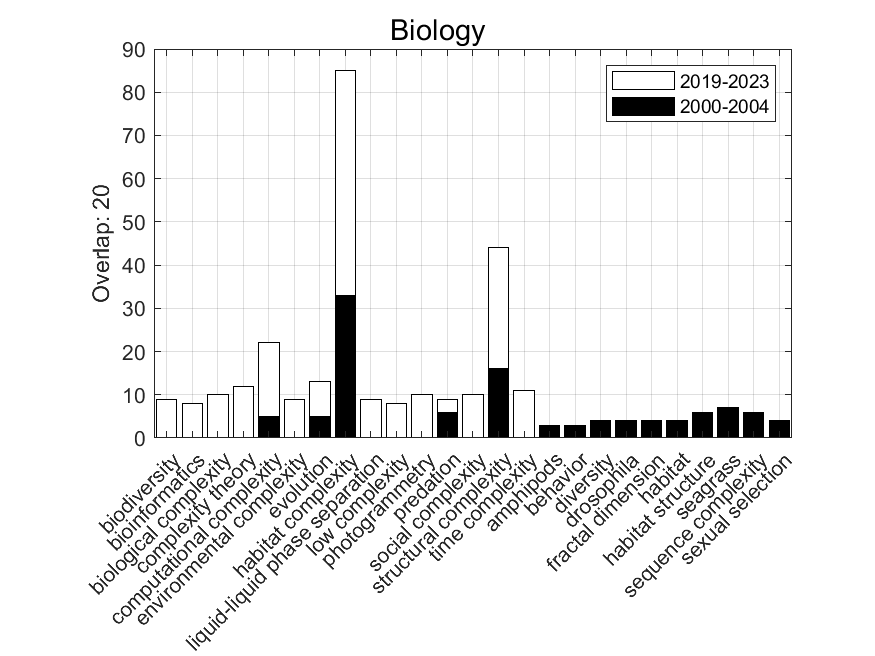}\includegraphics[width=.5\textwidth]{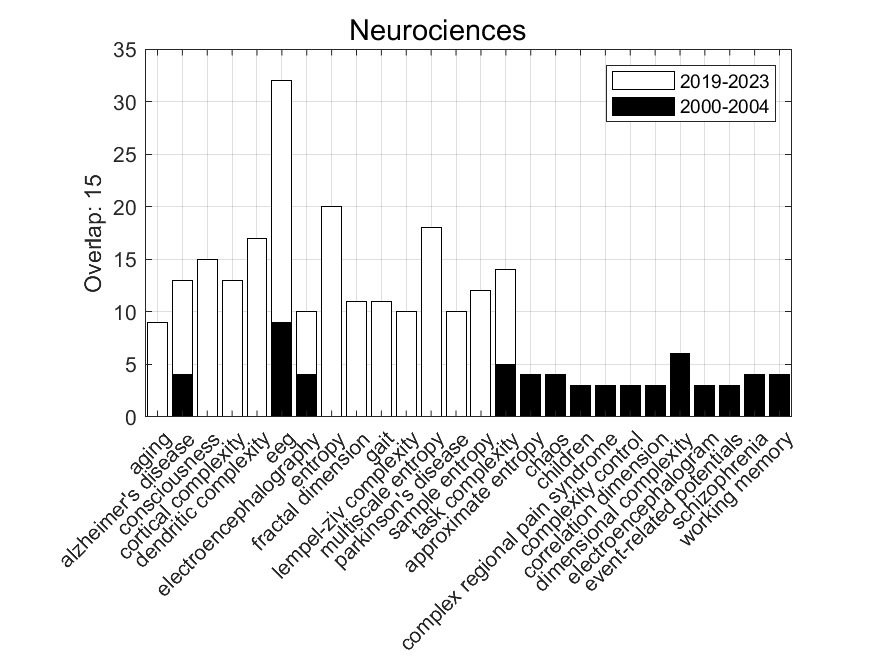}
\includegraphics[width=.5\textwidth]{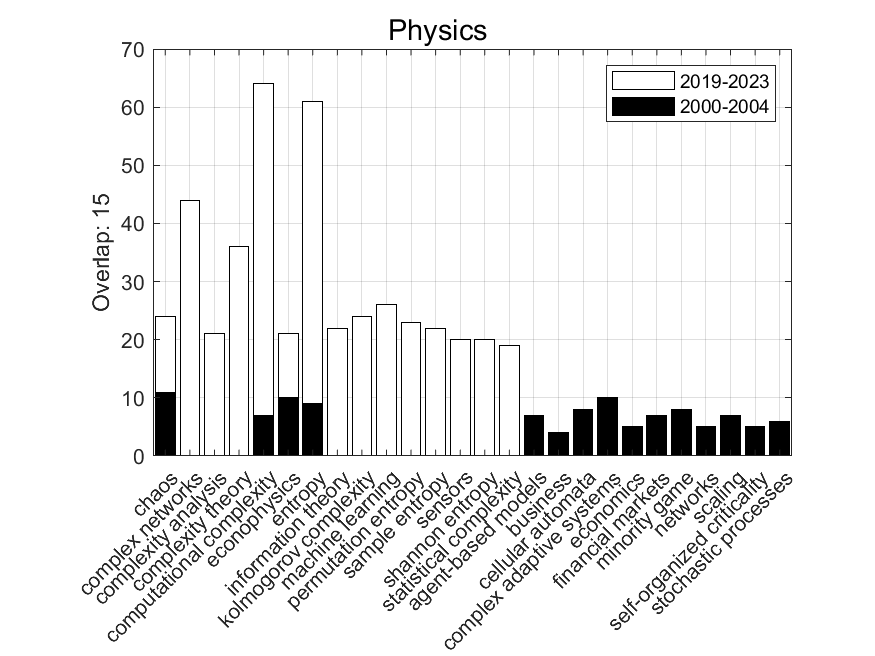}\includegraphics[width=.5\textwidth]{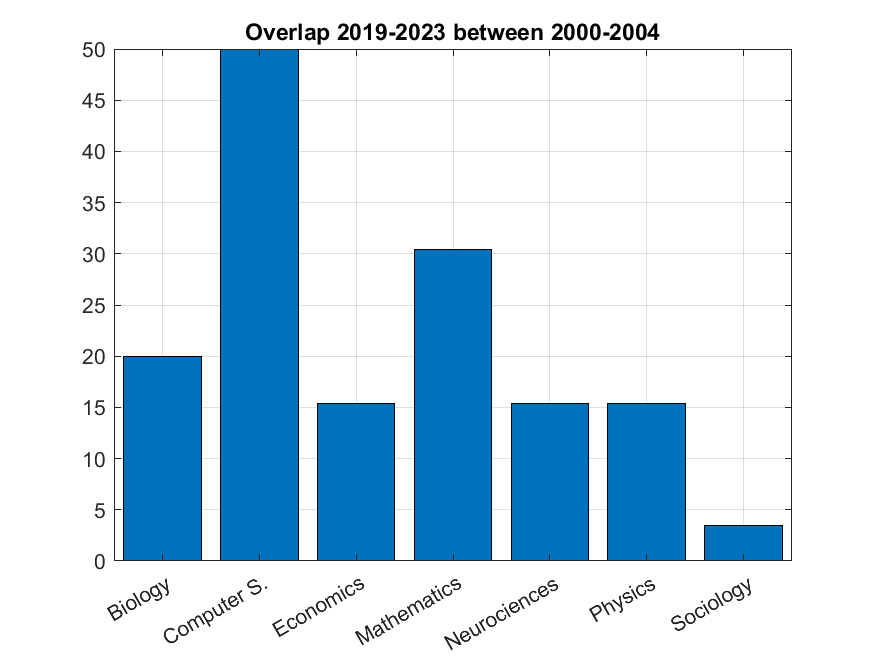}
  \caption{Author Keywords that co-occur with \texttt{Complexity}: in black in the first time interval (2000-2004) and colored white in the second one (2019-2023).}
  \label{fig4}
  \end{figure}
In Economics, there is great recent concern with Ecological trends, as shown by recent Author Keywords  like \texttt{Climate Change} and \texttt{Habitat Complexity} .
Economics is presented in the last histogram of Fig.\ref{fig3}. There, the most frequent co-occurring Author Keywords  are \texttt{Complex Adaptive Systems}, \texttt{Complexity Theory}, \texttt{Computational Complexity} and \texttt{Innovation}.
Physics is presented in Fig.\ref{fig4}. There, the most frequent co-occurring Author Keywords   are \texttt{Chaos}, \texttt{Computational Complexity}, \texttt{Econophysics} and \texttt{Entropy}.
In Biology, besides the leading and ubiquitous \texttt{Computational Complexity}, co-occurring Author Keywords  include \texttt{Evolution, Habitat Complexity, Predation} and \texttt{Structural Complexity}, as the first histogram in Fig.\ref{fig4} shows.
In Neurosciences, the time-overlapping occurring Author Keywords  consist of \texttt{Alzheimer's}, \texttt{EEG} and \texttt{Task Complexity}, showing that Author Keywords  are closely related to specific problems and/or instruments in the field.

The last histogram in Fig.\ref{fig4} shows the values of the Overlapping (O) coefficient between Author keywords in the two different time periods, for each research area. These areas display great heterogeneity in the amount of overlapping, with a huge difference between Computer Science and Sociology.

\section{Co-occurring Author Keywords  and keywords Plus®}

keywords Plus® are index terms automatically generated from the titles of the articles which are cited by each article in the sample. Keywords Plus terms must appear more than once in the bibliography and are ordered from multi-word phrases to single terms. Keywords Plus augments traditional keywords or title retrieval\footnote{http://www.garfield.library.upenn.edu/essays/v13p295y1990.pdf} .
Tab.\ref{tab2} shows a list of all multi-word phrases comprising the word \texttt{Complexity} and co-occurring with Author Keyword \texttt{Complexity} in Keywords Plus® in 2019-2023.

\begin{table}[h!]
    \centering
    \small{
    \begin{tabular}{cc}
    Multi-word phrases co-occurring with & Complexity in Keywords Plus®\\
    \hline
    Complexity Theory & Social Complexity\\
    Cognitive Complexity & Institutional Complexity\\
    Habitat Complexity & Task Complexity\\
    Economic Complexity & Dendritic Complexity\\
    Statistical Complexity & Kolmogorov Complexity\\
    Computational Complexity & Time Complexity\\
    Parameterized Complexity & Linear Complexity\\
    Topological Complexity & Polynomial Complexity\\
    Cortical Complexity & Complexity Analysis\\
    Structural Complexity & Low Complexity\\
    Environmental Complexity & Product Complexity\\
    Project Complexity & Time Complexity\\
    \hline
\end{tabular}}
\caption{Multi-word phrases co-occurring with \texttt{Complexity} in Keywords Plus in all research areas.}
\label{tab2}
\end{table}

\begin{figure}[h!]
  \centering
\includegraphics[width=.5\textwidth]{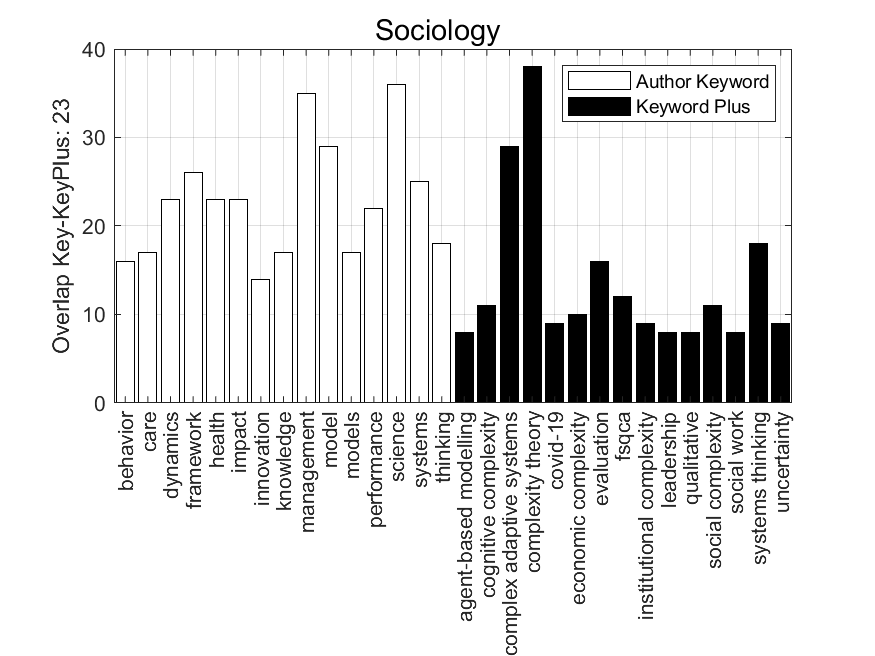}\includegraphics[width=.5\textwidth]{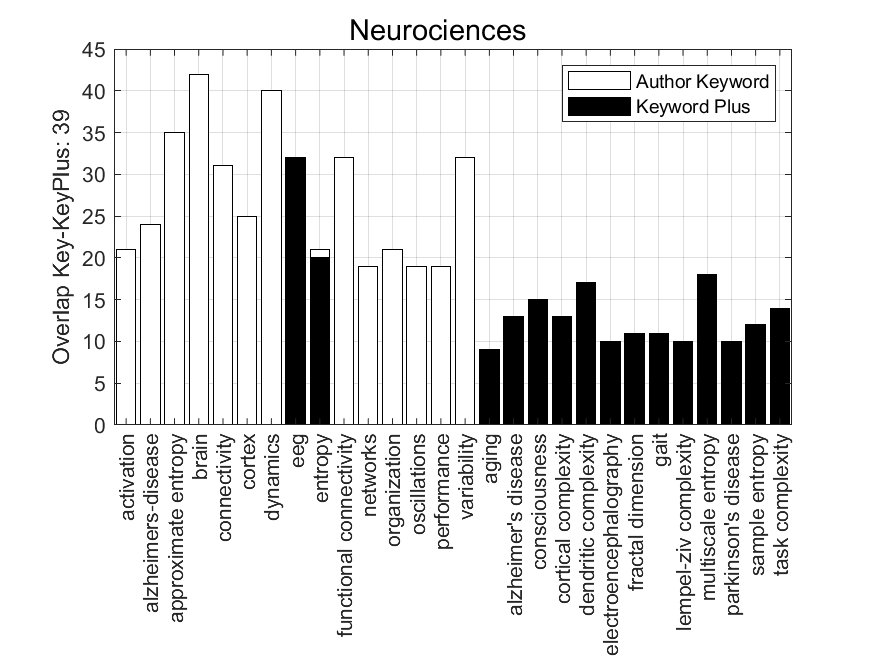}
\includegraphics[width=.5\textwidth]{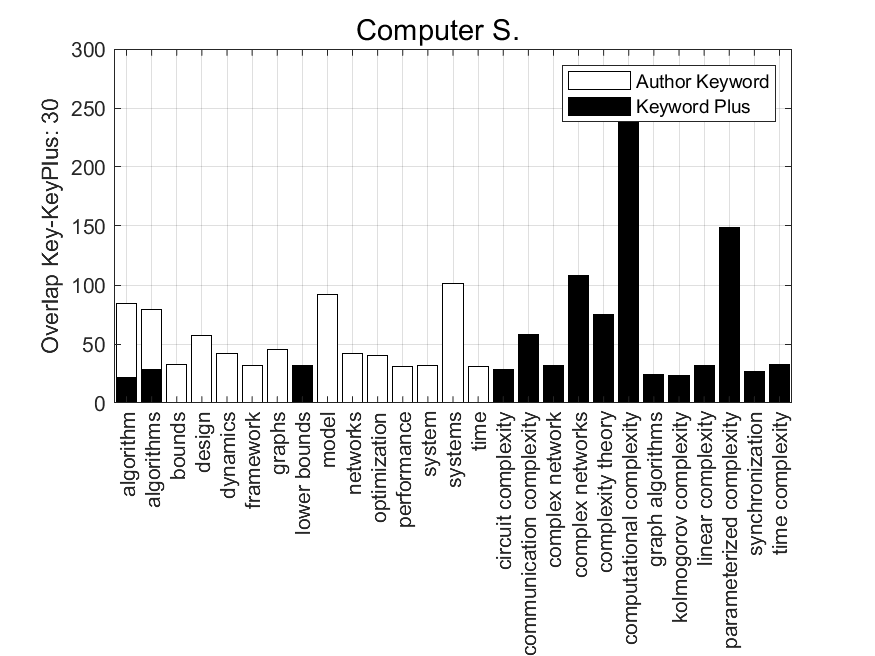}\includegraphics[width=.5\textwidth]{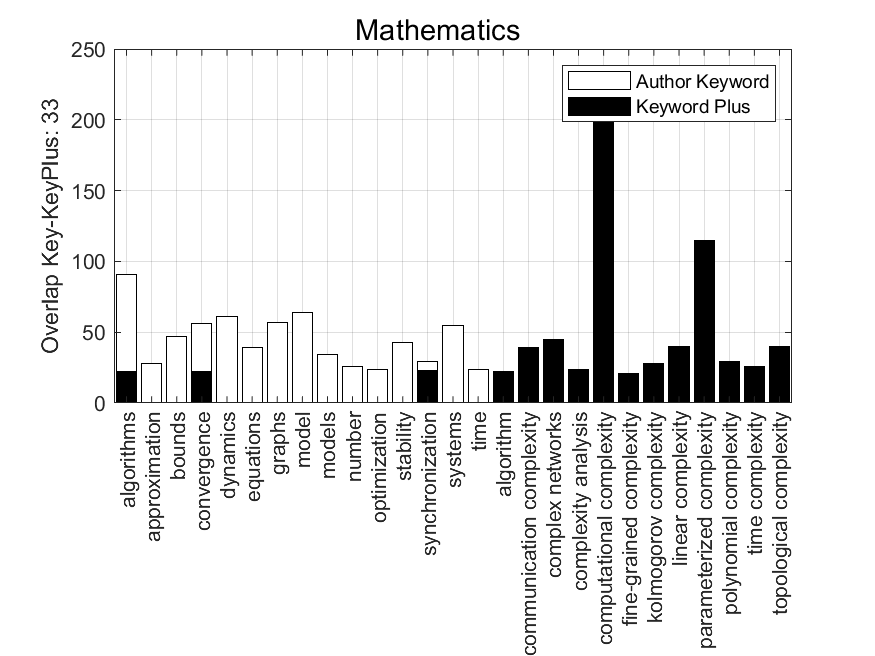}
\caption{Author Keywords  and keywords Plus that co-occur with \texttt{Complexity}.}
 \label{fig5}

\end{figure}

\begin{figure}[h!]
  \centering
\includegraphics[width=.5\textwidth]{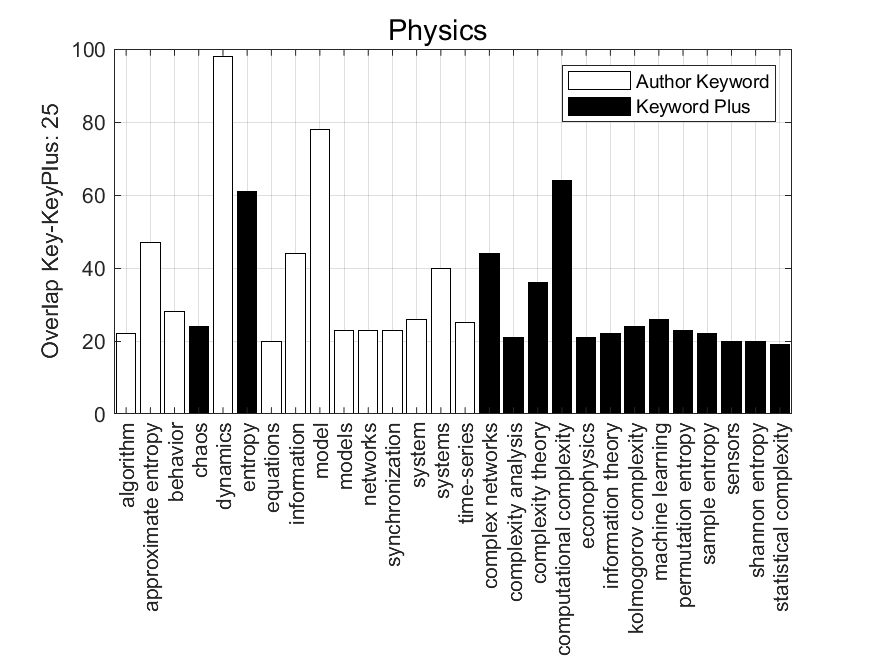}\includegraphics[width=.5\textwidth]{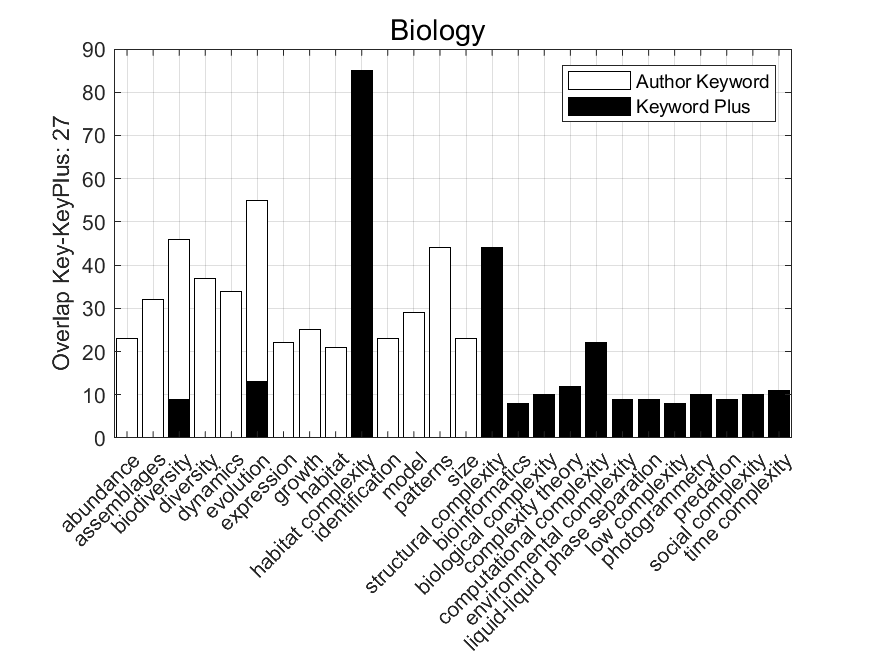}
\includegraphics[width=.5\textwidth]{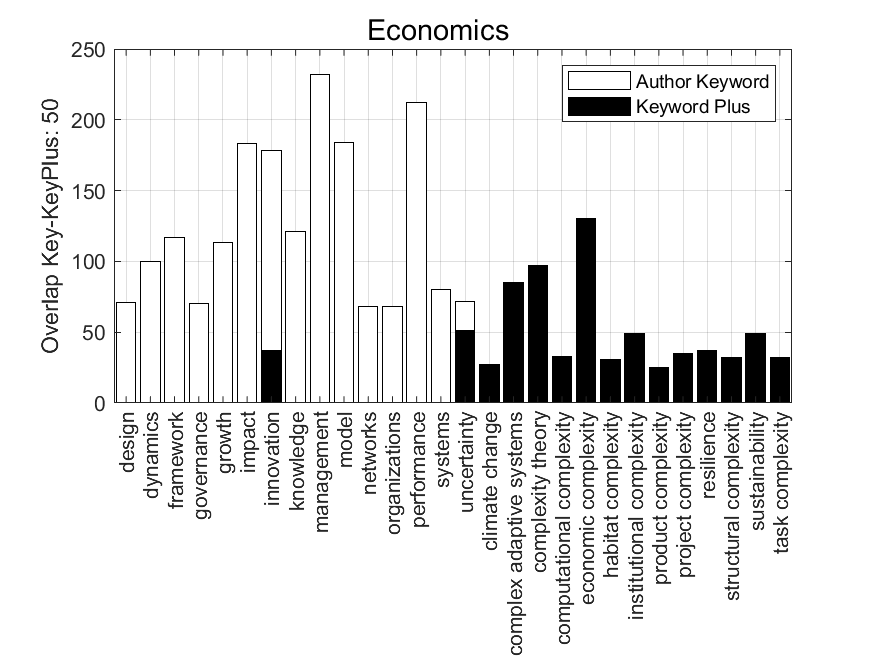}\includegraphics[width=.5\textwidth]{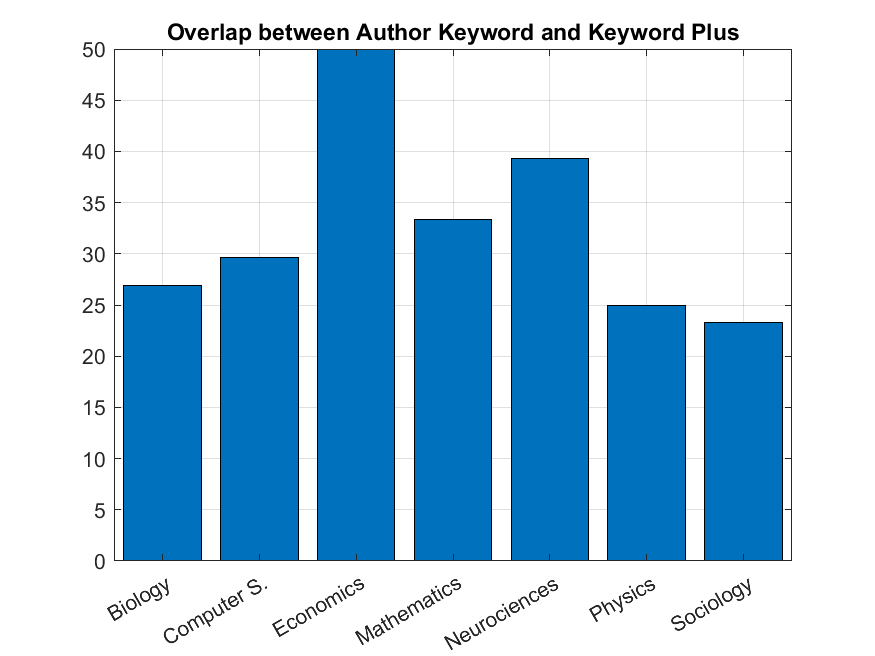}
\caption{Author Keywords  and keywords Plus that co-occur with \texttt{Complexity}.}
 \label{fig6}
\end{figure}

Figures \ref{fig5} and \ref{fig6} show the distributions of the frequency of Author Keywords and keywords Plus®, both co-occurring with Author Keyword \texttt{Complexity} in papers published in the time interval 2019-2023. The last histogram in Fig.\ref{fig6} shows the values of Overlapping (O) between Author Keywords and keywords Plus® computed for each research area. Unlike the values of Overlapping presented in the last histogram of \ref{fig4}, now, the overlap of the seven research areas shows a more homogeneous distribution.



 They do not highlight distinguishing features of the seven research areas. On the contrary, looking at the number of multi-word phrases co-occurring with  Complexity in Keywords Plus® along the seven research areas, one sees that areas as Economics, Computer Science and Mathematics display the highest number of those multi-word phrases (9, 10 and 11, respectively) while the opposite is observed in Sociology and Neurosciences (respectively, 5 and 4).

Because the multi-word phrases in Keywords Plus® are generated from the titles of cited articles and appear more than once in the bibliographies, the prevalence of a large number of multi-word phrases co-occurring with  Complexity in Keywords Plus® and comprising the word \texttt{Complexity} (as those listed in Table.\ref{tab2}) may be related to the existence of a greater consensus in the choice of terms in a given research area. It would, therefore, reflect the establishment, the setup and maturity of a discipline w.r.t. the application of Complexity Sciences.

Differently, research areas where the number of multi-word phrases co-occurring with \texttt{Complexity} in Keywords Plus® is small seem to display a less consensual practice regarding terms in the titles of cited articles and appearing more than once in their bibliographies.

In the next section, we consider the sets of Author Keywords co-occurring with \texttt{Complexity}  by research area and define for, each research area, the corresponding co-occurrence network of Author Keywords. Networks are defined by considering the 15 most frequent Author Keywords co-occurring with \texttt{Complexity} found in the seven research areas.
As in Section 5, the focus lies in the most recent time period (2019-2023), the list of Author Keywords co-occurring with \texttt{Complexity} by research area can be found in the x-axis of the histograms presented in  Fig.\ref{fig5} and Fig.\ref{fig6}.

\section{Networks}

The induction of a network is strongly dependent on the method by which networks are defined from a certain data set. The definition of the nodes and links connecting the elementary units of a system may occur in many different ways, therefore defining nodes and links depends strongly on the available empirical data and on the questions that a network analysis aims to address. Here, we address the hypothesis of the emergence of different topological structures of the networks of Author Keywords across the seven research areas being studied.

\subsection{Defining networks}

Bipartite networks are defined from the subsets of articles which may be related by co-occurring keywords that co-occur with \texttt{Complexity}.

The frequency of co-occurrence of each pair of keywords defines the existence of every link in the networks of Author Keywords. The resultant networks are therefore weighted graphs where nodes are keywords and the weight of each link corresponds to the frequency of co-occurrence of the linked pair of Author Keywords.

A bipartite network $N$ consists of two partitions of nodes $V$ and $W$, such that edges connect nodes from different partitions, but never those in
the same partition. A one-mode projection of such a bipartite network onto $V$ is a network consisting of the nodes in $V$; two nodes $v$ and $v\prime $
are connected in the one-mode projection, if and only if there exist a node $w\in W$ such that $(v,w)$ and $(v\prime, w)$ are edges in the corresponding
bipartite network ($N$). In the following, we explore bipartite networks of Author Keywords and their corresponding one-mode projections.

\subsection{Networks of Author-Keywords}

Each bipartite network $N^A$ of keywords consists of the following partitions:

\begin{itemize}
\item the set $S_A$ of the $n^A$ 15 most frequent Author Keywords found in the research area $A$ as presented in x-axis of the histograms in  Fig.\ref{fig5} and Fig.\ref{fig6}.

\item the set of articles ($P_{A}(t)$) $t=\{2019,...,2023\}$.
\end{itemize}

In each network ($N^A(t)$), two keywords are linked if and only if they co-occur in at least one article of $P^A(t)$. Naturally, the links in each network $(N^A(t))$ are weighted by the number of papers a pair of keywords share in $P^A(t)$.

Consequently, every link $L^A(i,j)  \in N^A$ takes value in the set $V=\{ 1,2,...,$ size $(P^A)\}$.\footnote{The index $t$ can be dropped since there is just one time interval under study.}

The induction of the networks ($N^A$) for each research area provides a dense representation of the relationships among the
Author Keywords co-occurring in that research area.
However, it so happens that the densely-connected nature of these networks does not help to characterize their topological structures. The large
number of links make the extraction of the truly relevant connections
forming the network a challenging problem. One first step in the direction
of extracting relevant information from each network is to
obtain its corresponding minimum-spanning-trees.

\subsubsection{minimum-spanning-trees}

In the construction of a Minimum spanning tree (MST) by the \textit{nearest neighbor} method, one
defines the Author Keywords as the nodes ($n_{i}$) of a weighted
and connected graph. From the $nxn$ distance matrix $D_{i,j}$, a hierarchical clustering is
performed using the \textit{nearest neighbor} method. Initially $n$ clusters
corresponding to the $n$ Author Keywords are considered. Then, at each step, two
clusters $c_{i}$ and $c_{j}$ are clumped into a single cluster if

\begin{center}
$d\{c_{i},c_{j}\}=\min \{d\{c_{i},c_{j}\}\}$
\end{center}

with the distance between clusters being defined by

\begin{center}
$d\{c_{i},c_{j}\}=\min \{d_{pq}\}$ with $p\in c_{i}$ and $q\in c_{j}$
\end{center}

This process is continued until there is a single cluster. This clustering
process is also known as the \textit{single link method}, this being the method
by which one obtains the MST of a graph (\cite{araujo2000}).
In a connected graph, the MST is a tree of $n-1$ edges that minimizes the
sum of the edge distances. In a network with $n$ nodes, the hierarchical
clustering process takes $n-1$ steps to be completed, and uses, at each
step, a particular distance $d_{i,j}$ $\in $ $D$ to clump two
clusters into a single one.

\begin{figure}[h!]
  \centering
\includegraphics[width=.5\textwidth]{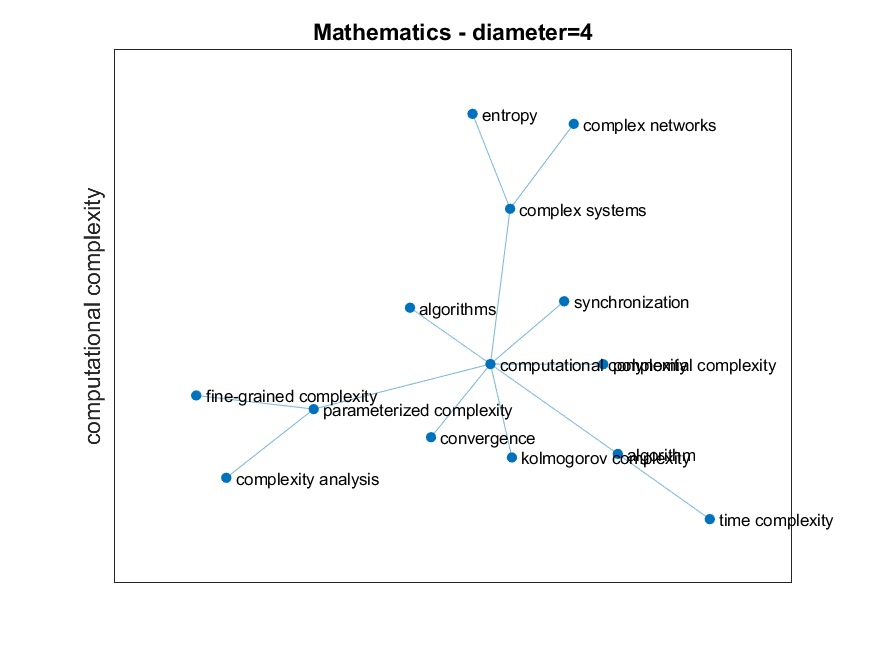}\includegraphics[width=.5\textwidth]{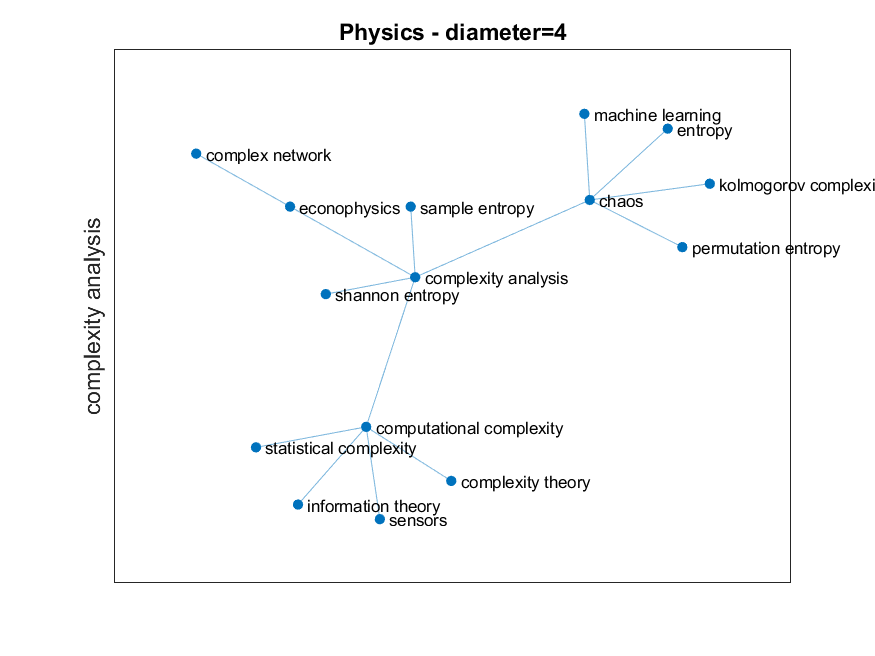}
\includegraphics[width=.5\textwidth]{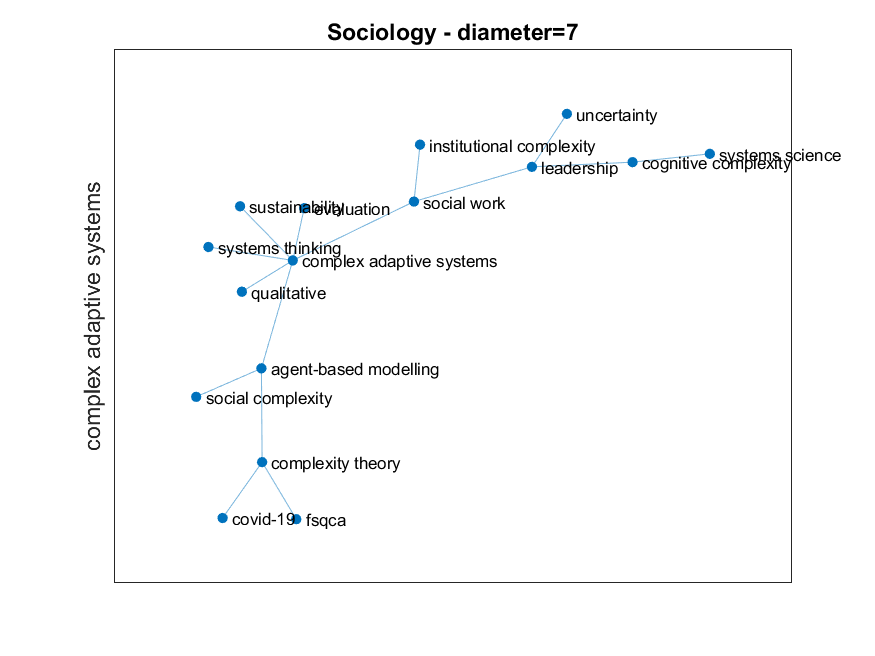}\includegraphics[width=.5\textwidth]{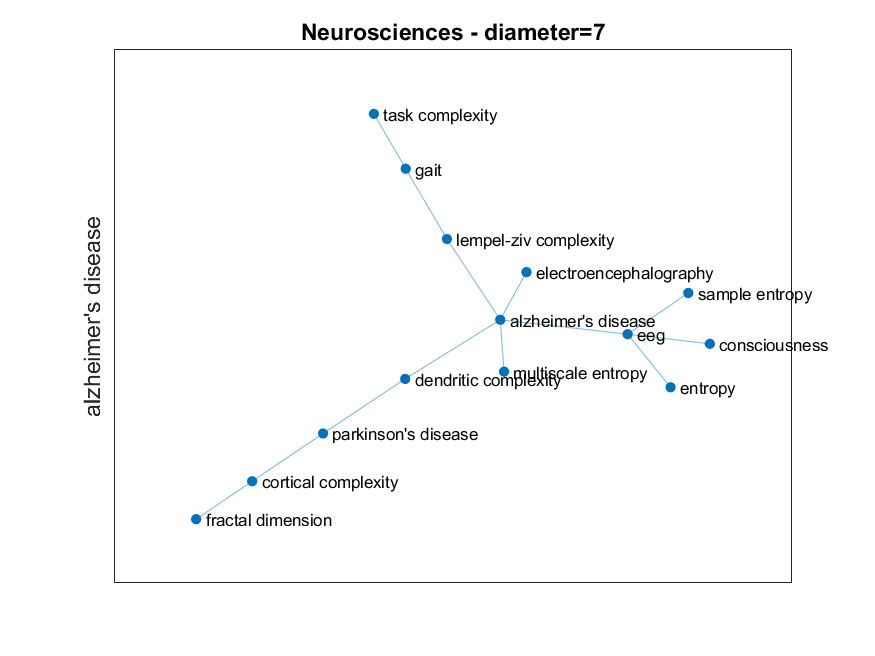}
\includegraphics[width=.75\textwidth]{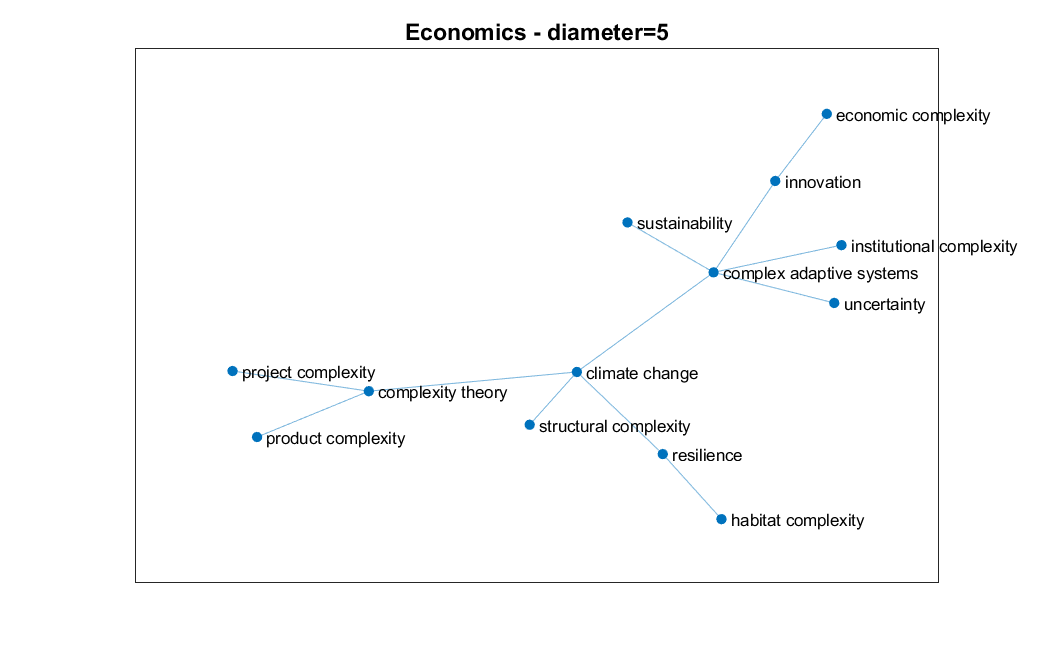}
\caption{The minimum-spanning-trees of the Networks of co-occurring Author Keywords that co-occur with \texttt{Complexity} in five research areas.}
 \label{fig7}
\end{figure}

The networks in Fig.\ref{fig7} show the MSTs of the co-occurring Author Keywords that co-occur with \texttt{Complexity} in five research areas: Mathematics, Physics, Sociology, Neurosciences and Economics.

The minimum-spanning-trees allow for the identification of three important coefficients that are not directly observed  in the original networks.

\begin{enumerate}
\item Branches and leaves: the way nodes organize themselves in different ramifications of the tree,

\item Diameter:  the longest path of all shortest paths between any two nodes in the MST.

\end{enumerate}

The first one is the number of branches and leafs $(b)$ and leaves $(l)$ in the MST, i.e., respectively, the number of
nodes with degree greater than one and number of
nodes with degree one. The second coefficient is the MST diameter ($d$),
measuring the shortest distance between the two most distant nodes on the
tree. The choice of these coefficients makes it possible to characterize tree motifs
according to their different shapes: from a pure \emph{star} to a pure \emph{path} motif.

It so happens that when the number of nodes of the tree is greater than 2,
and depending on the motif that the MST approaches, its diameter ranges in
between $2$ and $N-1$ ($2\leq d\leq N-1$). The closer $\frac{d}{N-1}$ is to
1, the lesser the similarity of the MST to a \emph{star} motif.
Moreover, the number of leafs ranges in between exactly the same values but
in the opposite direction: the closer $l$ is to 1, the lesser the
similarity of the MST to a \emph{path} motif.

The titles of the plots of each MST in Fig.\ref{fig7} comprise the values of the diameters of each MST. There, the foundational research areas in Complexity Sciences, such as Mathematics and Physics, display short diameters while research areas that started to apply Complexity tools more recently display much greater values of the diameter of their MST, as in the case of Neurosciences and Sociology. Their diameter is almost twice as large as those of Mathematics and Physics.

Small diameters indicate closer topological distances, showing that in the foundational research areas of Complexity the topological distances between co-occurring Author Keywords that co-occur with \texttt{Complexity} tend to be shorter, since some of their sub-networks approach a star motif, with one or two Author Keywords behaving like a hub to which several other keywords are connected.

The opposite situation characterizes Sociology and Neurosciences. There, each MST approaches a path motif, with larger topological distances between the pairs of co-occurring Author Keywords that co-occur with \texttt{Complexity}.
Economics relies in between those extremes, showing a diameter slightly greater than Mathematics and smaller than Sociology. Interestingly is the observation that, together with \texttt{Complex Adaptive Systems} the node with higher centrality is \texttt{Climate Change}.

The distances between keywords become larger as the number of weakly connected keywords increases. If, conversely, the corresponding MST approaches a star motif, the number of leafs increases and the corresponding diameter
decreases, as in the case of Mathematics and Physics. 
Tab.3 shows the topological coefficients computed from the MST of five research areas.

\begin{center}
\begin{tabular}{l|c|c|c|c|c}
\hline
\multicolumn{1}{c|}{\small MST} & {\small Math.} & {\small Physics} & {\small %
Sociology} & {\small Neuros.} & Economic \\ \hline
\multicolumn{1}{c|}{${\small N}$} & {\small 14} & {\small 15} &
\multicolumn{1}{|c|}{\small 16} & {\small 15} & {\small 13} \\ \hline
\multicolumn{1}{c|}{${\small d}$} & {\small 4} & {\small 4} &
\multicolumn{1}{|c|}{\small 7} & {\small 7} & {\small 5}\\ \hline
\multicolumn{1}{c|}{$l$} & {\small 10} & {\small 11} & \multicolumn{1}{|c|}%
{\small 10} & {\small 7}  & {\small 8}\\ \hline
\multicolumn{1}{c|}{$\frac{d}{N-1}$} & {\small 0.30} & {\small 0.26} &
\multicolumn{1}{|c|}{\small 0.44} & {\small 0.47} & {\small 0.41} \\ \hline
\label{tab3}
\end{tabular}

{\small Table 3: Topological coefficients from the MSTs of Mathematics, Physics, Sociology, Neurosciences and Economics}
\end{center}

Although the five networks are very similar in size, there is a remarkable
difference in the values obtained for their diameters. The MST of both Sociology and Neurosciences display
much larger diameters, showing that the distances among keywords are
large. In addition, the considerably smaller number of leaves ($l$) in Neurociences indicates
that this network exhibits an entirely different structure when compared with
the other research area. It seems that in areas where the use of the keyword \texttt{Complexity} started later, the existence of co-occurring keywords of strong centrality (hubs) is less frequent, which contributes to larger topological distances (and larger diameters) in the trees. By contrast, in foundational research areas such as Mathematics and Physics some highly frequent and very central co-occurring keywords, such as \texttt{Computational Complexity} and \texttt{Complex Adaptive Systems} , are hubs that cause the MST to approach a star-like motif.

Looking at the ratio $\frac{|d-l|}{N}$ across the different research areas helps to emphasize the distinguishing
structures of the MST that characterizes the networks of Sociology and Neurosciences.

It is worth noting that the proximity between two keywords on the MST depends on the connection strength (the
weight of the links) in each network of keywords, meaning that when two keywords
co-occur in many papers of a given research area (therefore being strongly
connected) they occupy close positions on the corresponding MST.

As previously observed, Author Keywords  co-occurring with \texttt{Complexity} in the later time period are more likely to belong to the specific research area under study. Now, looking at the trees in areas where the use of the keyword \texttt{Complexity} started later, one sees that the few keywords with higher centrality are also those describing specific concepts in the field, like \texttt{Alzheimer's} and \texttt{EEG} in Neurosciences.

\section{Concluding Remarks}

In the terms of the research questions raised at the beginning of this paper, we conclude that:

\begin{enumerate}
    \item There is a large increase in the number of papers from 2000-2004 to 2019-2023, and an even larger increase is observed in the number of papers having \texttt{Complexity} as an Author Keyword.
    Among the seven research areas being studied, Economics, together with Physics and Sociology display the highest increase.
    \item Author Keywords  co-occurring with \texttt{Complexity} in the earlier time period (2000-2004) tend to rely mostly on foundational concepts. Meanwhile, those co-occurring with \texttt{Complexity} in the later time period (2019-2023) are more likely to belong to a specific field of research.
    \item A large number of multi-word phrases co-occurring with  Complexity in Keywords Plus® and comprising the word \texttt{Complexity} seems to be related to the existence of a greater consensus in the choice of terms in a given research area, therefore reflecting its maturity regarding the application of Complexity Sciences.
    \item The MSTs of the foundational research areas in Complexity Sciences, such as Mathematics and Physics, display short diameters while research areas that started to apply Complexity tools more recently display much greater values of the diameter of their MST, as in the case of Neurosciences and Sociology.
    \item  In areas where the use of the keyword \texttt{Complexity} started later, the existence of co-occurring keywords of strong centrality (often specific concepts) is less frequent, contributing to larger topological distances. By contrast, in foundational research areas such as Mathematics and Physics some highly frequent and very central co-occurring keywords lead the corresponding MST to a small diameter while approaching a star-like motif.
    \item Such difference in the topological coefficients helps to emphasize the distinguishing structures that characterizes the networks of the seven research areas.
\end{enumerate}

\newpage
\textbf{{Acknowledgments}}

\vspace{0.5cm}

The authors acknowledge financial Support from FCT – Fundação para a Ciência e Tecnologia (Portugal). This article is part of the Strategic Project UIDB/05069/2020. The authors acknowledge financial Support from FCT – Fundação para a Ciência e Tecnologia (Portugal).\\

\section*{Declarations}

\begin{itemize}

\item Funding \\

This article is part of the Strategic Project UIDB/05069/2020. The authors acknowledge financial Support from FCT – Fundação para a Ciência e Tecnologia (Portugal).\\

\item Conflict of interest/Competing interests\\

The authors have no conflicts of interest to declare that are relevant to the content of this article.\\

\item Ethics approval: Not applicable \\

\item Consent to participate: Not applicable\\

\item Consent for publication: Not applicable\\

\item Availability of data and materials\\

Data is available at \\

\item Code availability\\

Code will be available at a GitHub public repository.\\

\item Authors' contributions\\

T. Araújo:  Methodology, Software, Supervision, Writing-Reviewing and Editing.\\
F. Louçã:  Methodology, Supervision, Writing-Reviewing and Editing.\\
A. Abreu:  Supervision, Writing-Reviewing and Editing.\\

\end{itemize}



\end{document}